\documentclass[fleqn,10pt]{wlscirep}

\usepackage{physics}
\usepackage{amssymb}
\usepackage{amsmath}
\usepackage{latexsym}
\usepackage{braket}

\usepackage{url}
\usepackage{xcolor}
\definecolor{newcolor}{rgb}{.8,.349,.1}

\usepackage[utf8]{inputenc}
\usepackage[T1]{fontenc}
\title{Methods for the construction of interacting many-body Hamiltonians with compact localized states in geometrically frustrated clusters}

\author[1,*]{F. D. R. Santos}
\author[1]{R. G. Dias}
\affil[1]{Departamento de F\'{i}sica $\&$ I3N, Universidade de Aveiro, Campus
	Universit\'{a}rio de Santiago,
	3810-193 Aveiro, Portugal.}

\affil[*]{filipedrsantos@ua.pt}

\begin{abstract}
Adding interactions to many-body Hamiltonians of geometrically frustrated lattices often leads to diminished subspaces of localized states. In this paper, we show how to construct interacting many-body Hamiltonians, starting from the non-interacting tight-binding Hamiltonians, that preserve or even expand these subspaces. The methods presented involve modifications in the one-body network representation of the many-body Hamiltonians which generate new interacting terms in these Hamiltonians.
The subspace of many-particle localized states can be preserved in the interacting Hamiltonian, by projecting the interacting terms onto the subspace of many-body extended states or by constructing the interacting Hamiltonian applying origami rules to the network.  
Expanded subspaces of localized states are found if interacting terms that mix subspaces with different number of particles are introduced.  
Furthermore, we present numerical methods for the determination of many-body localized states that allows one to address larger clusters and larger number of particles than those accessible by full diagonalization of the interacting Hamiltonian. These methods rely on the generalization of the concept of compact localized state in the network. Finally, we suggest a method to determine localized states that use a considerable fraction of the network.
\end{abstract}
\begin{document}

\flushbottom
\maketitle
\thispagestyle{empty}

\section{Introduction}
Flat-band systems have been heavily studied in the past two decades.\cite{huber2010bose,imada2000superconductivity,neupert2011fractional,peotta2015superfluidity,rhim2019classification,sun2011nearly,tang2011high,urban2011barrier,wang2011fractional} The attention gathered by these systems is due to the interesting phenomenon they manifest, such as fractional quantum Hall effect \cite{tsui1982two} or high-temperature superconductivity \cite{heikkila2016flat}, among others. Flat bands are associated to subspaces of one-body localized states, and such states are a consequence of destructive interference between different classical trajectories of particles and are associated with particular lattice geometry and particular choice of the hopping parameters of the respective tight-binding (TB) Hamiltonian (hopping terms between nearest neighbors in the simplest cases, but longer range hoppings can also allow the existence of localized states).\cite{Dias2015,Derzhko2015,tovmasyan2016effective}

\begin{figure}[!t]
	\centering
	\includegraphics[width=1 \textwidth]{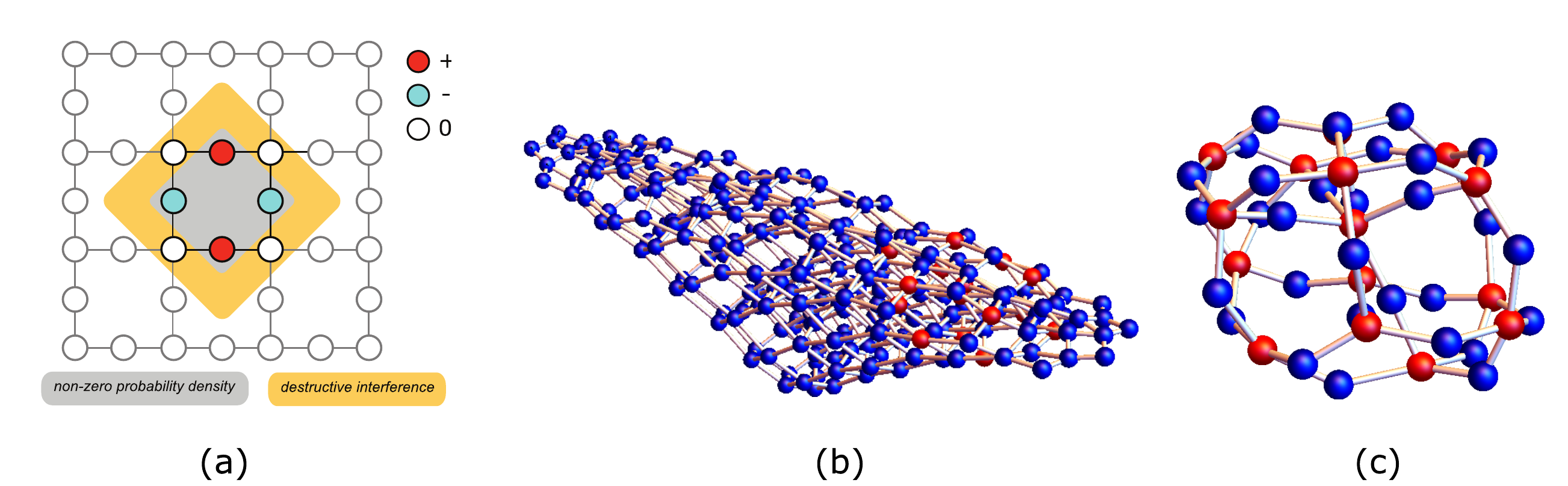}
	\protect\caption{(a) One-body CL state in a Lieb cluster. This state can be interpreted as a standing wave with normalized wavefunction amplitudes (+1,0,-1,0,+1,...) in a small region of the lattice (and zero in the rest of the lattice). (b) Network associated with the tight-binding Hamiltonian for a Lieb lattice with four plaquettes with two holes as particles (in a ferromagnetic background). The set of amplitudes given (red(blue) nodes have finite(zero) particle wavefunction amplitude) corresponds to a many-body localized state. This state is compact in the network and can still be interpreted as a standing wave. In (c), the same state is shown but the blue nodes that have no connections to red nodes were dropped. We designate this structure as a bubble. The compact many-boy localized state corresponds to the interlocking of one-body standing waves in the network. Figure generated in Mathematica 8, https://www.wolfram.com/.}
	\label{fig:LocState}
\end{figure}

From the flat-band Bloch eigenstates, applying the inverse Fourier transform, and properly choosing the relative phase between the flat-band Bloch states, it is possible to write maximally localized Wannier states.\cite{marzari2012maximally,marzari1997maximally} These Wannier states may be in certain lattices compact, but in general they will have small tails in the neighboring unit cells. A flat-band basis of compact localized (CL) states usually still exists in the latter case, if one drops the condition of an orthogonal basis. This is case of the Lieb lattice where the most compact one-body localized states are described by standing waves that travel along one plaquette, and the particle wavefunction amplitude is zero in the rest of the lattice \cite{Dias2015,Derzhko2015} [see Fig. \ref{fig:LocState}(a)].
The previous example of the Lieb lattice allows one to conclude that interacting many-body localized states may still exist in the presence of interacting terms that require overlap of the flat-band states if the number of interacting particles is less than the number of plaquettes and the overlap can be avoided. These are states that are insensitive to the interacting terms. In simple cases (for example in the case of the Lieb lattice or diamond chain), one-body CL states are described by standing waves with normalized amplitudes (+1,0,-1,0,+1,...) along a short path. However, in general, the flat-band subspace is perturbed by the introduction of interacting terms and it is not obvious if compact (many-body) localized states survive and how to determine them. In the case of interacting many-body localized states, the localized particles may be quite far apart in the lattice and the notions of standing wave and CL state needs to be generalized.

Recently, we have suggested a different approach to tackle this problem.\cite{santos2019hole} It is possible to describe many-body localized states through the network generated from the adjacency matrix associated with the respective Hamiltonian. In this representation, a many-body interacting Hamiltonian can be interpreted as a one-body Hamiltonian with possibly long range hoppings and local potentials in the many-body Wannier base, where the single particle moves in the network defined by the finite Hamiltonian matrix elements between the many-body Wannier states (nodes of the network) [see \cite{santos2019hole} for details]. In the network representation, localized many-body states may still be described by independent standing waves [see Fig. \ref{fig:LocState}(b)], the majority of them with support in a small compact region of the network. These compact many-body states correspond to interlocked one-body standing waves in the network, which we designate as bubbles [see Fig. \ref{fig:LocState}(c)].\cite{santos2019hole} The dimension of these bubbles is related to the average number of links per node (degree of the node) in the network, which is determined by the number of particles.

In this paper, we show how to construct interacting many-body Hamiltonians that allow the existence of CL states and expand some of the ideas that were introduced in  reference \cite{santos2019hole}. We present methods that involve modifications in the real lattice, in the network or directly in the Hamiltonian, such that the subspace of many-body localized states of the non-interacting Hamiltonian is \textbf{diminished, expanded or preserved}. 
In addition, since it becomes computational demanding to diagonalize many-body Hamiltonians and impractical to draw the corresponding networks (the matrices become too large even for a few number of particles), we suggest some methods that allow one to work with parts of a lattice or network, at a time, reducing the size of the matrices to diagonalize and the size of the networks to draw. We also present an algorithm that allows one to determine many-body localized states that are neither compact in the real lattice or in the network.
We apply these methods to determine full set of localized states of Lieb clusters and diamond chains.

\section{Preserving CL states: projecting interactions out of the subspace of localized states}

It is possible to add any kind of interacting terms to a Hamiltonian such that the subspace of states admits exactly \textbf{the same many-body localized states} (states such that all particles are localized) of the non-interacting Hamiltonian, as long as the interactions are projected out of the subspace of localized states (in other words, the interaction is projected onto the subspace of extended states). Since the subspace of localized  states is orthogonal to the subspace of extended states, one has $\hat{P}_{ext} \ket{loc} = 0$ and $\hat{P}_{loc} \ket{ext} = 0$, where $\hat{P}_{ext}$ and $\hat{P}_{loc}$ are projector operators that act only in, respectively, extended and localized states. Note that states $\ket{loc}$ have only localized particles and $\ket{ext}$ has at least one particle in an extended state (but one or more particles may be localized).
This is different from another possible approach where: (i) the single particle  annihilation and creation site operators are written as a linear combination of Wannier states of the set of bands of the non-interacting model; (ii) a projection of the interacting Hamiltonian onto extended states is obtained discarding in  the single particle  annihilation and creation site operators, the part in the linear combination corresponding to the Wannier states of the flat band (see \cite{tovmasyan2016effective,tovmasyan2013geometry} for an application of this method). The latter is a more severe projection method while the one we present here can be described as the  projection closest to the identity operator that applied to the interacting network recovers the many-body localized states of the  non-interacting network.

\begin{figure}[!ht]
	\centering
	\includegraphics[width=1 \textwidth]{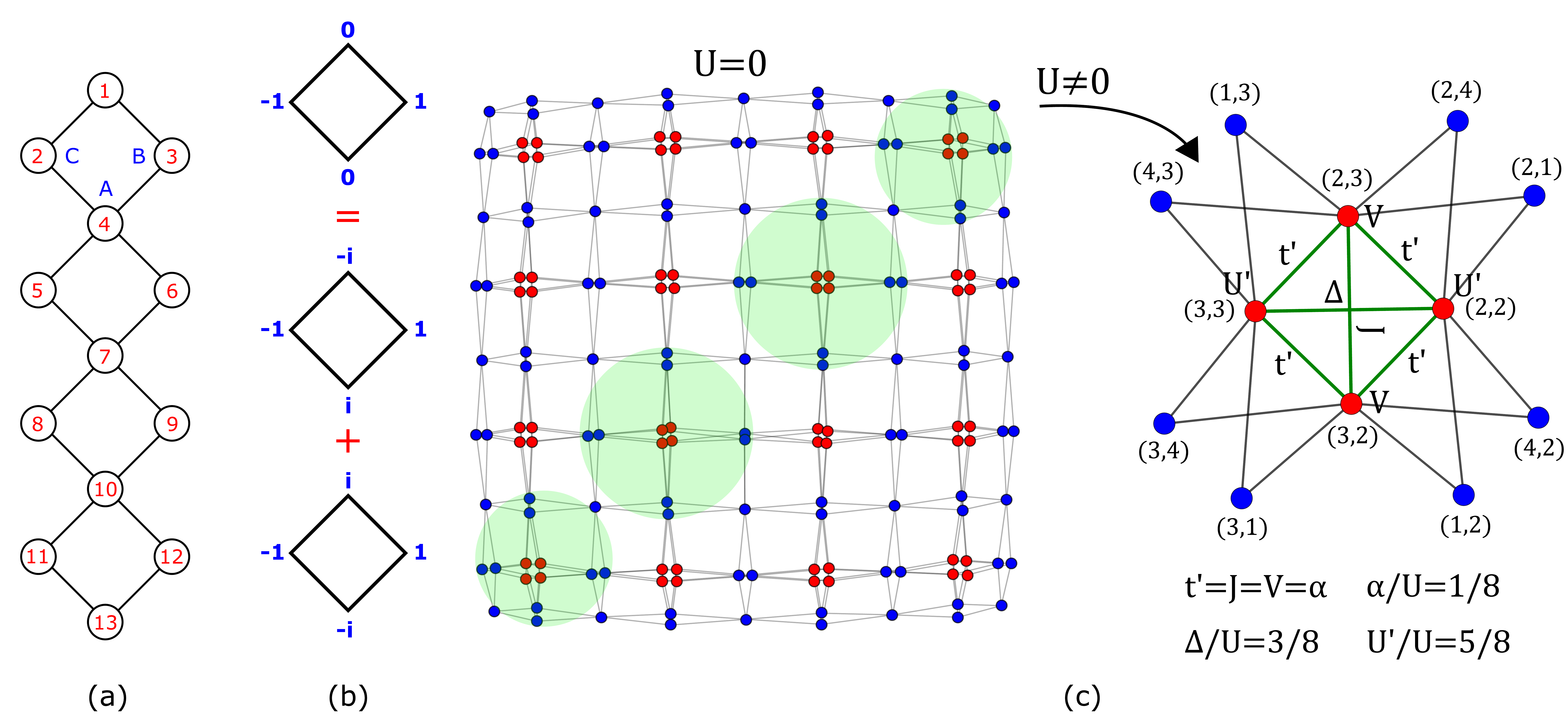}
	\protect\caption{(a) Diamond chain with four plaquettes. (b) One-particle localized state of a diamond plaquette. This state result from the combination of two one-plaquette states with opposite momentum. (c) Network associated with the TB Hamiltonian for a diamond chain with four plaquettes, with one spin up and one spin down. The state shown corresponds to a combination of all sixteen CL states, with the sites at red(blue) having finite(zero) particle wave-function amplitude. If Hubbard interacting terms are added and projected onto the subspace of extended states, the previous network is modified in the anti-diagonal squared bubble structures (marked by green circles). These modifications are illustrated in detail (at the right) for the top right squared bubble structure. The nodes that correspond to electronic configuration with the spins occupying sites B and C (sites 2 and 3 in this example) are linked to each other, and these links correspond to long range hoppings, $t'$, pair tunneling terms, $\Delta$, and spin exchange terms, $J$. The projection of the Hubbard terms also generate long range spin interactions, $V$, and the lowering of the Hubbard interaction in the red nodes, $U'<U$. Figure generated in Mathematica 8, https://www.wolfram.com/.}
	\label{fig:LocState3}
\end{figure}

Adding terms to the Hamiltonian implies modifications in the network, so how can the network change and still be associated to a Hamiltonian with the same localized states?
In order to answer this question, for simplicity, we discuss this method first in the case of the diamond chain, shown in Fig. \ref{fig:LocState3} (a). The diamond chain has three types of sites, A, B and C.
The tight-binding Hamiltonian for the diamond chain with spinless fermions, assuming zero on-site potentials and periodic boundary conditions, is given by
\begin{equation}\label{HTB}
\centering
\begin{split}
\centering
H_{\text{TB}}=t\sum_{j=1}^{N_c}\left[(A_{j}^{\dagger}B_{j}+A_{j}^{\dagger}C_{j}+\text{H. c.})+(A_{j}^{\dagger}B_{j+1}+ A_{j}^{\dagger}C_{j+1}+ \text{H. c.})\right],
\end{split}
\end{equation}
where the index $j$ labels the unit cells, $N_c$ is the number of unit cells, and $t$ is the hopping constant. The first term corresponds to intra-unit cell hoppings terms and the second term corresponds to inter-unit cell hopping terms. 
From the diagonalization of $H_{\text{TB}}$, one obtains three electronic energy bands. Two of those bands are dispersive (non-flat) with energy
$
\epsilon_{k,\pm} =\pm 2t \sqrt{2} \cos (\sqrt{2}k_x/2).
$
The third band is a zero energy (dispersionless) flat-band and it is $N_c$-fold degenerated. The one-particle CL states associated with the flat band arise from the anti-bonding combination of B and C orbitals,
$
\ket{loc; j} = (B_{j}^{\dagger}-C_{j}^{\dagger})/\sqrt{2}\ket{vac}.
$
These states result from the linear combination of the doubly degenerate opposite momentum Block states $k=\pi/2$ and $k=-\pi/2$ of a diamond plaquette [see Fig. \ref{fig:LocState3} (b)]. The flat band has only support in the B and C sublattices, and admits $N_{c}$ orthogonal localized states (the orthogonality of the localized states subspace basis makes the determination of the projection operators easier).

In Fig. \ref{fig:LocState3} (c), it is shown the network associated to the TB Hamiltonian of a diamond chain with four plaquettes, with one spin up and one spin down. Each node of the network corresponds to a position pair $(i_{\uparrow},j_{\downarrow})$ where $i_{\uparrow}$ is the position of the spin-up and $j_{\downarrow}$ is the position of the spin-down according to the site indexing shown in Fig. \ref{fig:LocState3} (a). The Hamiltonian admits sixteen two-particle localized states where each particle is in a localized state (dropping the spin index) $\ket{loc; j}$. These two-particle localized states in the network are linearly independent standing waves that form squared bubble structures (with twelve sites), with finite particle wave-function amplitude in the (red) nodes that correspond to having both spin up and spin down localized at sites B or C.
Adding interacting terms to the previous Hamiltonian may lead to a loss of localized states. For instance, in the presence of Hubbard interactions, the states in which the spin-up and the spin-down were localized in the same plaquette become extended (note that the Hubbard interaction becomes a local potential in the one-body network of many-particle Wannier states).

Nonetheless, if the Hubbard terms ($U \sum_i n_{i \uparrow} n_{i \downarrow}$, with $n_{i,\sigma}= X_{i, \sigma}^{\dagger} X_{i, \sigma}$, $X=A$, $B$ or $C$, and where the index $i$ labels the sites of the lattice) are projected onto the subspace of extended states, one finds that the interacting Hamiltonian admits exactly \textbf{the same number of localized states} as the non-interacting version of this Hamiltonian (we choose $U/t = 10$, but one could choose any finite value for $U$). 
In this case, the network is slightly different from the network shown in Fig. \ref{fig:LocState3} (c), having some modifications in the off-diagonal bubble like structures. The new network includes links that connect the red nodes in Fig. \ref{fig:LocState3}(c) that correspond to having both spin occupying sites B or C of the same plaquette.
These modifications are shown in Fig. \ref{fig:LocState3} (c) at the right, for the top right corner of the network (in the real lattice this correspond to having the spins occupying the top plaquette of the diamond chain shown in Fig. \ref{fig:LocState3} (a)). The new edges (drawn in green) correspond to new terms: long range hoppings, $t'$, [edges linking (2,3) to (2,2) to (3,2) to (3,3) to (2,3)], a pair tunneling term, $\Delta$, [edge from (2,2) to (3,3)] and a spin exchange term, $J$, [edge from (3,2) to (2,3)]. Besides these modifications, the new Hamiltonian includes long range spin interactions, $V$, between spins located in sites B and C of the same plaquette. Note also the lowering of the Hubbard interaction values for sites B and C, $U'<U$ (sites 2 and 3 in this example). The fractions $\alpha/U$ (where $\alpha= t'= J = V$), $\Delta/U$ and $U'/U$ are equal to, respectively, $1/8$, $3/8$ and $5/8$, and we found that these ratios remain the same for any value of $U$.

This means that it is possible to construct a Hamiltonian that combines different kinds of interactions in such a way that the effect of the Hubbard interacting terms is not felt by the subspace of CL states. Note that superconducting interactions and exchange interactions are often assumed to be competing interactions \cite{berk1966effect,gor1964ferromagnetism}, but in this case the different interactions are cooperating in order to generate the many-particle CL state subspace.
The interacting Hamiltonian can be written, symbolically, as a combination of such interactions in the following way
\begin{equation}\label{interactions}
\begin{split}
\hat{H}=\hat{t}+\hat{U}=\hat{t}+(\hat{U'}+\hat{V}+\hat{t'}+\hat{\Delta}+\hat{J})+(\hat{U}-\hat{U'})-(\hat{V}+\hat{t'}+\hat{\Delta}+\hat{J}).
\end{split}
\end{equation}
The term ($\hat{U'}+\hat{V}+\hat{t'}+\hat{\Delta}+\hat{J})$ does not affect the subspace of localized states and thus, the last two terms of Eq.\ref{interactions} can be seen as perturbation terms in what concerns the subspace of localized states.

\begin{figure}[!ht]
	\centering
	\includegraphics[width=1 \textwidth]{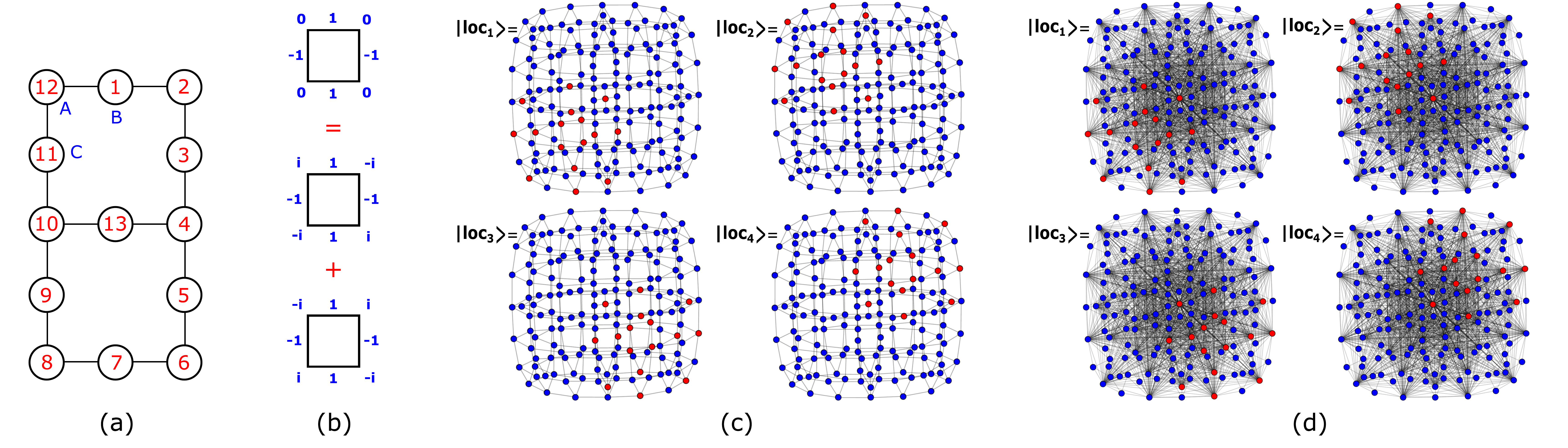}
	\protect\caption{(a) Lieb lattice with two plaquettes. (b) One-particle localized state of a Lieb plaquette. This state result from the combination of two one-plaquette states with opposite momentum. (c) Compact localized states of the TB Hamiltonian for the lattice in (a), with one spin up and one spin down, in the network representation. The standing-waves associated to the CL states share some nodes (with finite particle wavefunction amplitude), evidencing the non-orthogonal nature of the CL states subspace basis. (d) Network associated to the TB Hamiltonian with Hubbard interacting terms projected onto the subspace of extended states, for the lattice in (a), with one spin up and one spin down. All sites with finite particle wavefunction amplitude (red sites) are connected to each other, meaning that hoppings of particles or pair of particles (or displaced pairs of particles) between plaquettes are allowed. Figure generated in Mathematica 8, https://www.wolfram.com/.}
	\label{fig:LocState4}
\end{figure}

Let us now illustrate this method in more detail in the case of the Lieb lattice.
The Lieb lattice is a square bipartite lattice characterized by a unit cell with three types of sites: A (at the vertices), B (at the vertical edges) and C (at the horizontal edges) [see Fig. \ref{fig:LocState4} (a)].
The tight-binding Hamiltonian for spinless fermions in the Lieb lattice, assuming zero on-site potentials, is given by
\begin{equation}\label{HTB}
\begin{split}
H_{\text{TB}}=t\sum_{x=1}^{L_x}\sum_{y=1}^{L_y}\left[(A_{x,y}^{\dagger}B_{x,y}+A_{x,y}^{\dagger}C_{x,y}+\text{H. c.})+(A_{x,y}^{\dagger}B_{x,y-1}+ A_{x,y}^{\dagger}C_{x-1,y}+ \text{H. c.})\right],
\end{split}
\end{equation}
where periodic boundary conditions are implicit and where $L_x$ ($L_y$) correspond to the number of unit cells in the $x$ ($y$) direction, and $t$ is again the hopping parameter. The first and second terms correspond, respectively, to intra-unit cell and inter-unit cell hopping terms.
From the diagonalization of $H_{\text{TB}}$, one obtains two dispersive (non-flat) bands and one (dispersionless) flat band. The energy of the dispersive bands is given by
$
\epsilon_{k,\pm} =\pm 2t \sqrt{\cos^2 (k_x/2)+\cos^2(k_y/2)},
$
where $k_{\alpha}=2\pi n_{\alpha}/L_{\alpha}$ with $n_{\alpha}=0,1,...,L_{\alpha}-1$ and ${\alpha} \in {x,y}$.
The flat band is a zero energy band $L_x L_y$-fold degenerated, and the one-particle localized states associated with this band can be written as
$
\ket{loc; x,y} = (B_{x,y}^{\dagger}-C_{x,y}^{\dagger}+B_{x+1,y}^{\dagger}-C_{x,y+1}^{\dagger})/2\ket{vac}.
$
The flat band has only support in the B and C sublattices, or, in other words, flat band localized states have zero amplitudes at A sites.
The most compact one-particle localized states of Lieb lattices are localized in a single plaquette [see Fig. \ref{fig:LocState}], and are described by standing waves that results from the linear combination of doubly degenerate opposite momentum Bloch states $k=\pi/2$ and $k=-\pi/2$ of a single plaquette [see Fig. \ref{fig:LocState4} (b)].
Two side-by-side CL states overlap at one site (site at the middle of the rung). 
This implies that the set of one-particle maximally CL states generates a non-orthogonal basis of the flat band subspace, but they are still linearly independent, and thus the number of one-hole localized states is equal to the number of plaquettes, $N_{plaq}$. 
A global many-body basis of the Lieb lattice can be written as
$
\Big\{\ket{loc, j}  \Big\} \cup  \Big\{ \ket{ext,i} \Big\} 
$
where $\Big\{ \ket{ext,i} \Big\}$ is the orthogonal basis of many-particle extended states subspace  (again, recall that some of the particles in these states may be localized) and $\Big\{\ket{loc,j}  \Big\}$ is the non-orthogonal basis of compact many-particle localized states subspace.

Due to the fact that the CL states subspace basis, $\Big\{\ket{loc,j}  \Big\}$, is non-orthogonal, it follows that $\hat{P}_{loc} \neq \sum_{j} \ket{loc,j}  \bra{loc,j}  $, and instead the projector operator is given by (see Supplementary Material and Ref. \cite{soriano2014theory} for more details)
\begin{align}
\begin{split}
\hat{P}_{loc} &= \sum_{m,i} \ket{m} S_{m, i}^{-1} \bra{i},
\\
\hat{P}_{loc}^{\dagger} &= \sum_{m,i} \ket{i} S_{i, m}^{-1} \bra{m},
\end{split}
\end{align}
where $m$ runs over $\Big\{\ket{loc,m} \Big\}$, and $i$ runs over 
the global many-body basis of the Lieb lattice. 
In the case of a non-orthogonal basis set $\{i\}$ the overlap matrix obeys the condition $S_{ij}=\braket{i|j}\neq \delta_{ij}$ for some $i$ and $j$. If the basis $\big\{\ket{i} \big\}$ is normalized, i.e., $\braket{i|i}=1$ $\forall i$, then $S$ is symmetric (in our case $S$ will also be real since we work with real Hamiltonians). The matrix $S$ can be broken down into blocks in the following way
\begin{align}
\hat{\mathrm{S}}=\quad
\begin{pmatrix}
\hat{S}_{loc} & 0 \\
0 & \hat{I}_{ext}
\end{pmatrix},
\quad
\end{align}
where the block $\hat{I}_{ext}$ corresponds to the overlap between extended states, and, assuming an orthogonal basis for the subspace of extended states, it is just an identity matrix. The matrix elements of the anti-diagonal blocks are equal to zero due to the subspace of localized states being orthogonal to the subspace of extended states. The block $\hat{S}_{loc}$ refers to the overlap between localized states.

The four localized states of the TB Hamiltonian for the Lieb lattice with two plaquettes with one spin up and one spin down (with $t_{ij}=t$ $\forall ij$) are shown in Fig. \ref{fig:LocState4} (c).
The overlap matrix in the basis 
$
\Big\{\ket{loc, j}  \Big\} \cup  \Big\{ \ket{ext,i} \Big\} 
$ 
is in this case
\begin{align}
\hat{\mathrm{S}}=\quad
\begin{pmatrix}
\begin{matrix}
1 & -1/4 & -1/4 & 1/16 \\
-1/4 & 1 & 1/16 & -1/4 \\ 
-1/4 & 1/16 & 1 & -1/4 \\ 
1/16 & -1/4 & -1/4 & 1
\end{matrix}
& 0 \\
0 & \hat{I}_{ext}
\end{pmatrix}.
\quad
\end{align}
This overlap matrix is easy to understand looking at the states in Fig. \ref{fig:LocState4} (c). The values displayed in the 4x4 block ($\hat{S}_{loc}$) were obtained doing the inner product between the four localized (non-orthogonal) eigenstates, of the non-interacting Hamiltonian. That is, this block is $S_{ij}^{loc}=\braket{loc_i|loc_j}$, with $i,j=1,2,3,4$. Note that the particular order chosen for the states in Fig. \ref{fig:LocState4} (c) is irrelevant. Since these states have 16 equal finite absolute amplitudes in the network (red nodes), for each shared red node between two states of Fig. \ref{fig:LocState4} (c), one has a 1/16 contribution to the overlap (all contributions have the same sign). Thus, the overlap is 1, 1/4 and 1/16 when the localized states share respectively 16, 4 and 1 nodes of finite amplitude.

In Fig. \ref{fig:LocState4} (d), it is shown the network associated to the interacting Hamiltonian, 
\begin{equation}\label{HU}
\begin{split}
H_{\text{U}}=\sum_{\langle i j \rangle, \sigma} t_{ij} X_{i \sigma}^{\dagger} X_{j \sigma} + U \hat{P}_{ext} \sum_i n_{i \uparrow} n_{i \downarrow} \hat{P}_{ext}
\end{split}
\end{equation}
with Hubbard terms projected  onto the subspace of extended states,
where one introduced the spin-$1/2$ variable $\sigma$ (and the respective sum over the spin degrees of freedom in the Hamiltonian), $X=A$, $B$, or $C$, and $n_{i \sigma}=X^\dagger_{i \sigma} X_{i \sigma}$ and the sum over nearest neighbors $\langle i j \rangle$ for the Lieb lattice is given in Eq.~\ref{HTB}. The  matrix representation of the projection operator $\hat{P}_{ext}$ in the many-body Wannier basis can be determined numerically faster from $\hat{P}_{ext}=\hat{I}- \hat{P}_{loc}$ (where the matrix representation of $\hat{I}$ is the identity matrix and $\hat{P}_{loc}$ is given above) since in general the dimension of the subspace of localized states is much smaller than the dimension of the subspace of extended states.
This network is very different from the network associated with the non-interacting Hamiltonian, having new edges connecting all (red) nodes corresponding to electronic configurations with the spin particles occupying sites B and C, including nodes that correspond to having spins trapped in different plaquettes. This means that jumps of particles (or pair of particles) between plaquettes are allowed, and this is a consequence of the non-orthogonal nature of the CL states subspace basis. 
Despite these modifications, the new network is associated with an interacting Hamiltonian that holds exactly the same localized states of the Hamiltonian that lead to the network in Fig. \ref{fig:LocState4} (c), and the reasoning for this is the same as in the case of the diamond chain.
The terms that result from the projection of the Hubbard terms combine and negate the effect of the Hubbard interaction. In this case, two new kind of terms emerge from the projection of the interacting Hubbard terms: long range hopping terms that allow particles to jump between sites of different plaquettes, and an interaction that promotes tunneling of displaced pairs of spins to other B or C sites of the lattice.

\section{Preserving CL states: Origami rules in the network}\label{section3}

There is a complete set of modifications that can be applied to a geometrically frustrated  lattice that preserve the single-particle CL states. 
This set of rules \cite{Dias2015} has been proposed for non-interacting geometrically frustrated lattices and clusters, but they remain valid for many-body interacting systems, and better, these rules can be applied to the network generated from the adjacency matrix associated with a  many-body non-interacting Hamiltonian and the respective modifications of the network structure lead to modifications in the Hamiltonian, in the form of interactions. As an example, one of the Origami rules states that one may add or drop links between nodes of the network with zero many-body wavefunction amplitude and the resulting state is still a CL eigenstate of the modified many-body Hamiltonian.
Thus, it is possible to add interacting terms to a non-interacting Hamiltonian that admits compact many-body localized eigenstates, in such a way that the localized states of the non-interacting Hamiltonian are preserved.

\begin{figure}[!t]
	\centering
	\includegraphics[width=0.8 \textwidth]{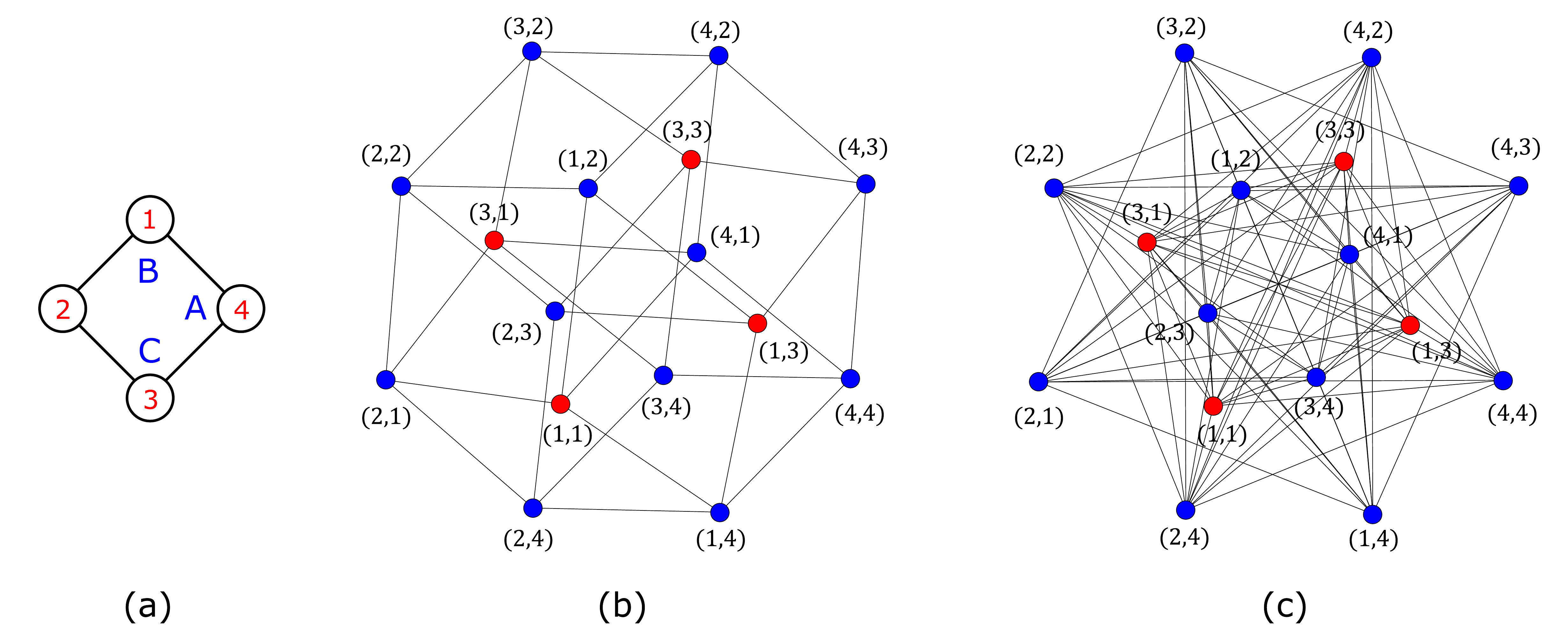}
	\protect\caption{(a) Diamond plaquette. (b) Network representation of the single localized state admitted by the TB Hamiltonian and (c) by the interacting Hamiltonain (with interactions that allow hoppings between all nodes of the network) for the lattice in (a) with one spin-up and one spin-down particles. Despite the differences between the two networks, the respective Hamiltonians admit the same localized state. Figure generated in Mathematica 8, https://www.wolfram.com/.}
	\label{fig:cd}
\end{figure}
In Fig.~\ref{fig:cd} (b),  it is shown the single localized state, in the network representation, admitted by the TB Hamiltonian for a single diamond plaquette (Fig.~\ref{fig:cd} (a)) with one spin up and one spin down particles, assuming that any additional plaquettes will  connect to sites 2 and 4.
As an example of application of the Origami Rules, one may introduce a term  $\sum_{ijmn} c_{i, \uparrow}^ \dagger c_{j, \downarrow}^ \dagger c_{m, \uparrow} c_{n, \downarrow}$ in the TB Hamiltonian, where the indices run over all sites of the network region shown in Fig.~\ref{fig:cd} (b). This term correspond to the first Origami rule that says that an additional set of hopping bonds may be introduced if they do not lower the local symmetry of the network region where the localized state has finite amplitudes and in this case the CL state is preserved. In the network, the term $c_{i, \uparrow}^ \dagger c_{j, \downarrow}^ \dagger c_{m, \uparrow} c_{n, \downarrow}$ becomes "hopping" terms between two nodes of the network. This term links all sites equally, keeping the symmetry of the network [see Fig.~\ref{fig:cd}(c)] and it is equivalent to a $c_{k=0,\uparrow}^ \dagger c_{k=0, \downarrow}^ \dagger c_{k=0, \uparrow} c_{k=0, \downarrow}$ term (which is a density operator product in the $k$-space of one plaquette). The CL state of the plaquette remains the same since it is a linear combination of $k=\pi/2$ and $k=-\pi/2$ Bloch states in the diamond plaquette. We opted to chose $k=0$ but other $k$ values are possible as long as they are different from $k=\pi/2$ and $k=-\pi/2$ (this is also valid for diamond chains with more plaquettes, and since all localized states can be written as a linear combination of CL states, one may use any linear combination of $k$ values of different plaquettes too). 

\section{Diminishing the number of CL states: adding interactions}

It it possible to add interacting terms to an Hamiltonian such that the new Hamiltonian still preserves some of the localized states of the original one, lifting partially the degeneracy of the subspace of localized eigenstates. For instance, as discussed previously in this paper, adding a Hubbard term to a TB Hamiltonian of a Lieb lattice (or diamond chain) with one spin up and one spin down particles leads to a diminished subspace of localized states. Note that, if the number of repulsive spins particles is larger than the number of plaquettes then no localized states survives.

However note that adding interactions may also have the opposite effect, i.e, it is possible to add interactions to a Hamiltonian that, originally, does not admit localized states and find localized states, or expand the subspace of localized states of a Hamiltonian that admits localized states in the first place. For the latter case, as example, we show in the next section  that the subspace of localized states of the TB Hamiltonian of diamond chains with spin up particles can be expanded by adding terms to the Hamiltonian that mix states that involve components with different spin up particle number.

\section{Generating new CL states: mixed filling  states}\label{section5}
\begin{figure}[!t]
	\centering
	\includegraphics[width=0.8 \textwidth]{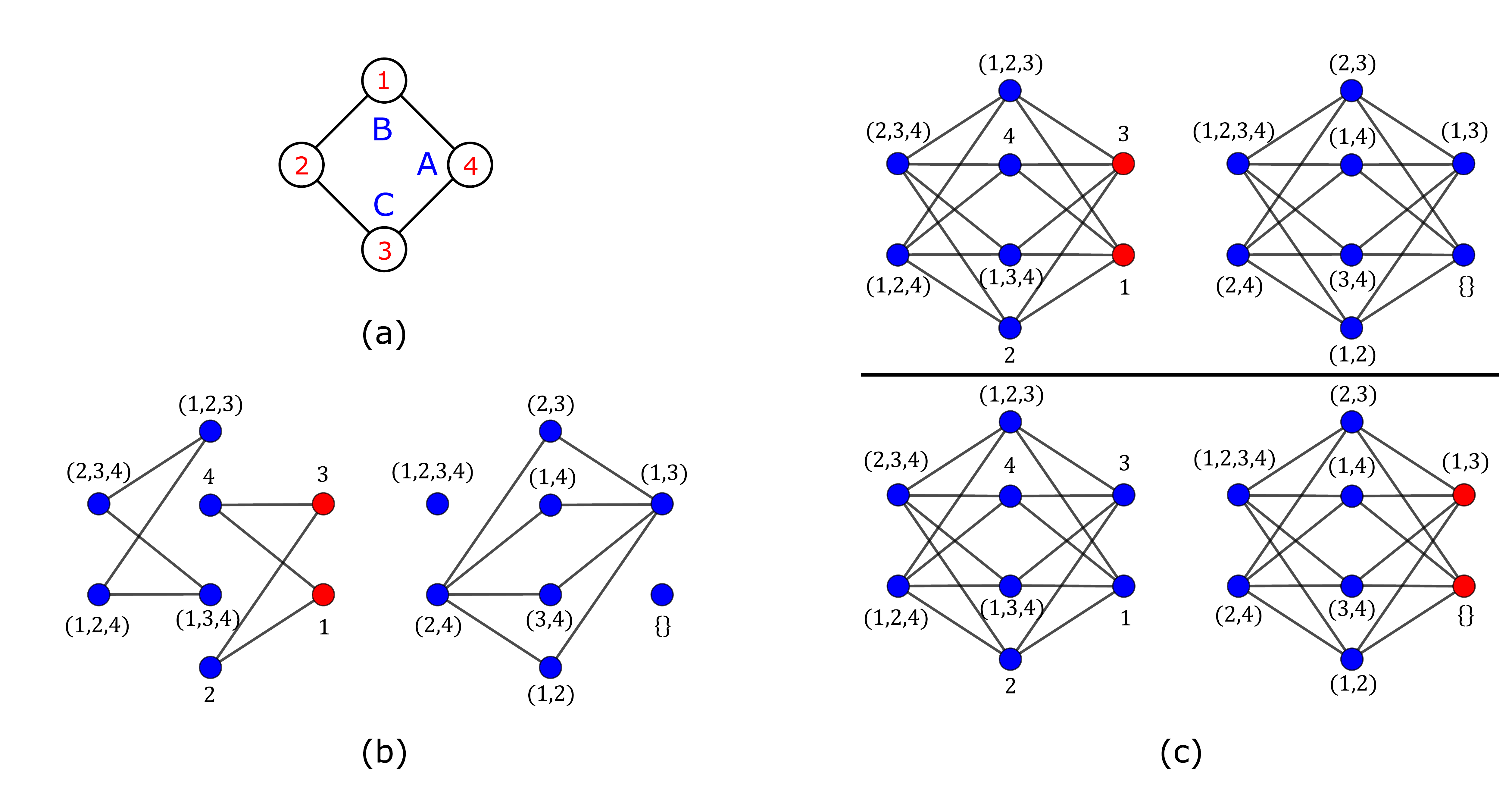}
	\protect\caption{(a) Diamond plaquette. (b) Networks associated to the non-interacting TB Hamiltonian and  (c) to the TB Hamiltonian with pair tunneling interacting terms of a single diamond plaquette for all different fillings of spin up particles. One localized state is shown (finite amplitudes in red) in the non-interacting network and two in the interacting network. Figure generated in Mathematica 8, https://www.wolfram.com/.}
	\label{fig:diamond1x1}
\end{figure}

One can generalize the argument given in section \ref{section3} to other types of interactions that do not involve the linear combination of $k=\pi/2$ and $k=-\pi/2$ plaquette Bloch states corresponding to the CL state of the diamond chain. 
For example, one can construct Hamiltonians that admit localized states that involve components with different particle number (with links connecting the nodes of the Hamiltonian networks of different fillings).
We illustrate this in the case of diamond plaquettes with different number of spin-up particles.
In Fig. \ref{fig:diamond1x1} (b), the network associated to the tight-binding Hamiltonian of a single diamond plaquette (Fig. \ref{fig:diamond1x1} (a)) for all possible fillings of spin-up particles is shown. The finite amplitudes of the single CL state in the network admitted by this Hamiltonian are shown in red. This state can be described as a one-body standing wave in the network, with finite particle amplitude in the nodes that correspond to having a single particle occupying the sites $1$ and $3$ (this agrees with the fact that, due to Pauli exclusion, localized states with more than one spin up particles in a single plaquette can not exist). Again we assume that additional plaquettes will be connected to sites 2 and 4.

In order to construct many-body CL states that involve different particle number we added the interacting pair tunneling terms $c_1 c_2 + c_1 c_4 + c_2 c_ 3+ c_4 c_3 + H.c.$ to the non-interacting Hamiltonian (recall that the CL state in the diamond plaquette is the anti-bonding combination of the $1$ and $3$ Wannier states). In Fig.~\ref{fig:diamond1x1} (c), the network associated with the new Hamiltonian is shown, as well as the amplitudes of the two CL states that are now found. 
In the left part of the network Wannier states of one and three particles are linked, while in the right part of the network Wannier states of four, two and zero particles are connected.
In the localized state shown at the top, the many-body particle wavefunction amplitude is finite for the single-particle Wannier states $1$ and $3$ (just as in the case of the localized state of the original, non-interacting Hamiltonian).
In the bottom localized state of Fig.~\ref{fig:diamond1x1}(c), the many-body particle wavefunction amplitude is finite for the node $(1,3)$ which corresponds to having two spin-up particles occupying the $1$ and $3$  sites of the real lattice and for the node corresponding to the vacuum state.
Longer range Hubbard-like type interactions may also lead to new localized states in a spinless diamond chain (such localized states  have been previously described in \cite{lopes2011interacting} and are due to the interplay of geometric frustration and interacting terms).

\begin{figure}[!ht]
	\centering
	\includegraphics[width=1 \textwidth]{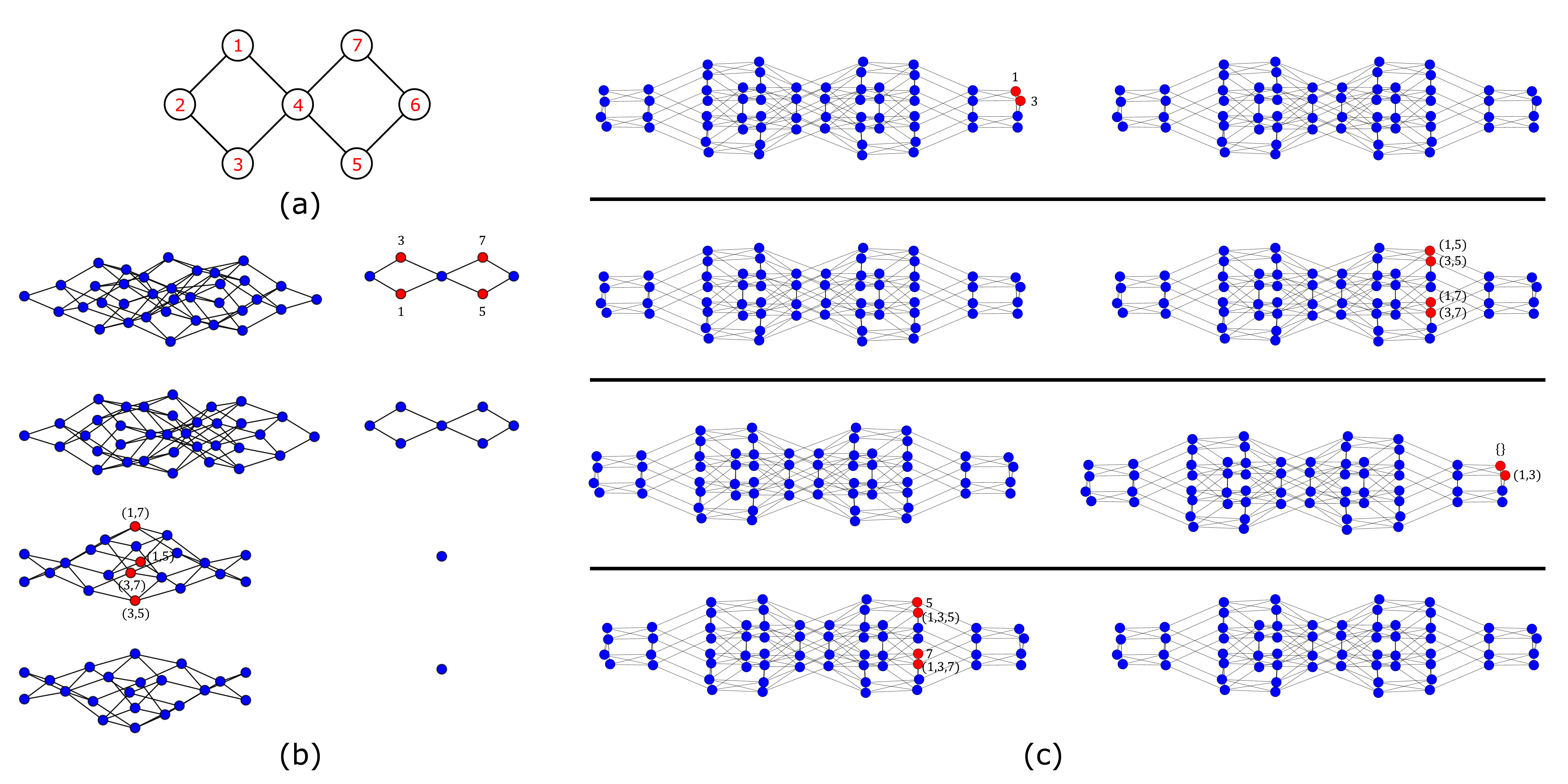}
	\protect\caption{(a) Diamond chain with two plaquettes. (b)  Three CL states in the network associated with the non-interacting TB Hamiltonian and  (c) four CL states [two are the same as those in (b)] in the network associated with the TB Hamiltonian with pair tunneling interacting terms, for the lattice in (a), for all different fillings of spin up particles. Figure generated in Mathematica 8, https://www.wolfram.com/.}
	\label{fig:diamond1x2}
\end{figure}

This method can be generalized for lattices with any number of plaquettes  (and in particular, for any geometrically frustrated lattice). In Fig.~\ref{fig:diamond1x2} (b), the network associated with the non-interacting Hamiltonian in the case of a diamond chain with two plaquettes [see Fig.~\ref{fig:diamond1x2} (a)] is shown, as well as the amplitudes of the three CL states. There are two one-body CL states, each corresponding to a standing wave traveling along one of the plaquettes, and a two-particle CL state with each spin-up particle occupying a different plaquette.
In Fig.~\ref{fig:diamond1x2} (c), the network for the same lattice is shown but now for the interacting version of the TB Hamiltonian where interacting terms $c_1 c_2 + c_1 c_4 + c_2 c_3 + c_4 c_3 + H.c.$ were introduced (these terms act only in the left plaquette and we dropped the spin index). The interacting Hamiltonian admits now four localized states and they involve components with different number of particles.
The first localized state occurs in the network with Wannier states of one and three particles (recall that the interaction acts only in the left plaquette). A single spin-up is localized in the left plaquette, having finite amplitude in the nodes $1$ and $3$.
The second localized state occurs in the network that links Wannier states of two and four particles and the two-particle localized state has finite amplitudes in nodes that correspond to having the spins localized at different plaquettes.
The third localized state involves a network of  Wannier states of zero, two and four particles, with finite amplitude in nodes that correspond to having both spins in the left plaquette.
The last localized state appears in the network of  Wannier states of one and three particles, with finite amplitude in the nodes $5$ and $7$ which correspond to having a single spin located in the right plaquette and in the nodes ($1,3,5$) and ($1,3,7$) which correspond to having two spins in the left plaquette and one in the right plaquette.

We emphasize that the above mixed-filling compact many-body states are found only if the  interacting term acts on a single plaquette.
If the interacting term acts in all lattice sites, this leads to a reduction of the number of localized states because the interaction leads to weight transfer between plaquettes, generating an extended version of some CL states.

\section{Finding many-particle CL states}\label{section6}
In the previous sections, we have described how to construct interacting Hamiltonians that admit many-particle CL states. The form of these states was known since it was the same as that of  the non-interacting Hamiltonian or because the Hamiltonian was defined in such a way that a particular CL eigenstate was present.
Below (with the exception of subsection \ref{eqamplitudes}), we discuss how to find many-particle CL states of an interacting Hamiltonian without any prior knowledge about its possible form or existence.
\subsection{Selecting parts of the network}

For large number of particles in a lattice, the diagonalization of the interacting  Hamiltonian in order to obtain the full set of localized states becomes numerically demanding. However, since most many-body localized states have support in a small region of the network, it is possible to determine the majority of the localized states by diagonalizing the Hamiltonian submatrix associated with that small region of the network.\cite{santos2019hole}

\begin{figure}[!h]
	\centering
	\includegraphics[width=0.7 \textwidth]{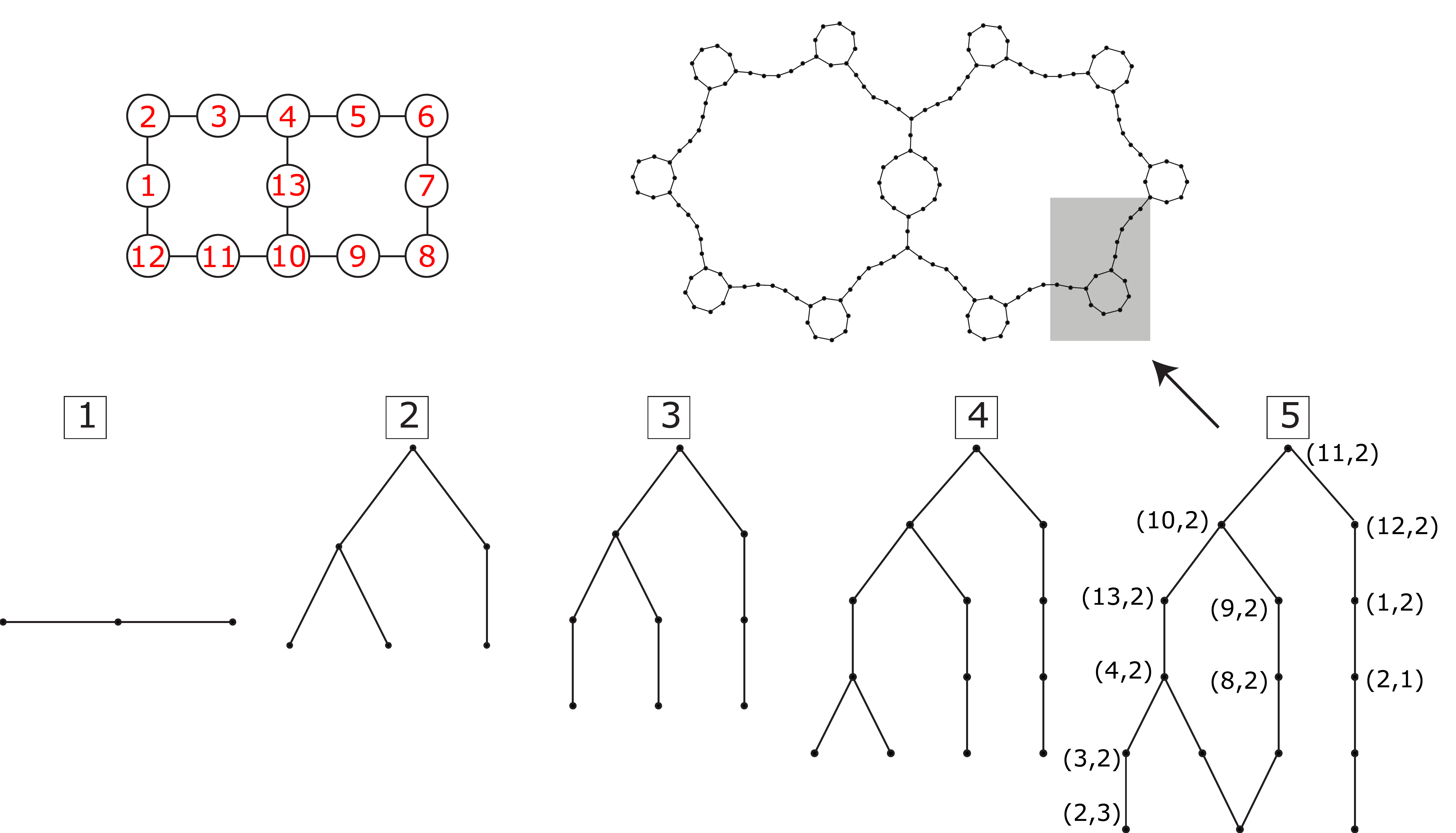}
	\protect\caption{Bottom: growth of the network (fully represented at the top right) for the TB Hamiltonian of a Lieb lattice with two plaquettes (top left) with one-hole and one flipped spin in a ferromagnetic background. These diagrams were obtained from a numerical method used to determine CL states. The method works in the following way: starting from an arbitrary node, the network is grown (diagrams 1-4) until a bubble-like structure is revealed (diagram 5), which indicates de presence of a CL states. This cycle is repeated, restarting  from another arbitrary node (not present in the cluster grown so far) until all CL states are determined. Figure generated in Mathematica 8, https://www.wolfram.com/.
	}
	\label{fig:grow}
\end{figure}

In Fig. \ref{fig:grow}, at the bottom, it is shown one cycle of our method to draw parts of the network and identify CL states. The diagrams 1-5 in Fig. \ref{fig:grow} sketch the growth of a small part of the network associated with the Lieb-$1 \times 2$ cluster with one hole and one flipped spin in a ferromagnetic background. Each node of the network corresponds to a position pair $(i_{\circ},j_{\downarrow})$ where $i_{\circ}$ is the position of the hole and $j_{\downarrow}$ is the position of the flipped spin according to the site indexing shown in Fig. \ref{fig:grow}. The method works in the following way. First, one arbitrarily chooses one node of the network (a state of the many-body Wannier basis) from which one starts to construct the network, by adding nodes with links to that initial node according to the elements of the adjacency matrix (see diagram 1). Next, one adds nodes with links to the current cluster of nodes and repeat this process (diagrams 2,3,4) until a bubble-like structure forms in the network (diagram 5), which may indicate the existence of a localized state (in order to verify if the bubble is associated to a localized state, one should determine the eigenstates of the Hamiltonian for this small region, and check if any of the states corresponds to a standing wave with zero amplitudes at the nodes that connect the bubble to the rest of the network). At the end of each step, one checks if any bubble-like structure emerge in the network and if it does, one diagonalizes the partial Hamiltonian. After such structure is found, one selects (randomly) another node, not present in the cluster grown so far, and starts to grow another part of the network, repeating the process until all bubble-like structures are grown.

Besides allowing to obtain localized states without having to diagonalize the full Hamiltonian, working with smaller adjacency matrices also always gives CL states while the diagonalization of the full network can generate "extended" versions of the localized states.
Note that some many-body localized states correspond to standing waves that use large portions of the network, and thus can not be obtained by using this method. Further below in this section we discuss how to determine those states.

\subsection{Drop sites in the lattice}

Another way to work with small matrices is to drop sites of the real lattice. Again, one-body (and some many-body) CL states are described by standing waves that travel only along a small part of the whole lattice, and so it is possible to drop one or a set of sites and still determine the localized states that have support in the remaining lattice. 
For instance, in the case of one-hole localized states of a Lieb ladder (or diamond chain), one can remove any number of outer-most plaquettes and still find CL states of the Hamiltonian associated with the original lattice. If one determines the subset of localized states for each set of dropped points, it is possible to obtain the full set of one-body localized states.
This method can be also applied to the many-body network, but it is  trickier to choose which nodes to drop. 

\begin{figure}[!h]
	\centering
	\includegraphics[width=0.9 \textwidth]{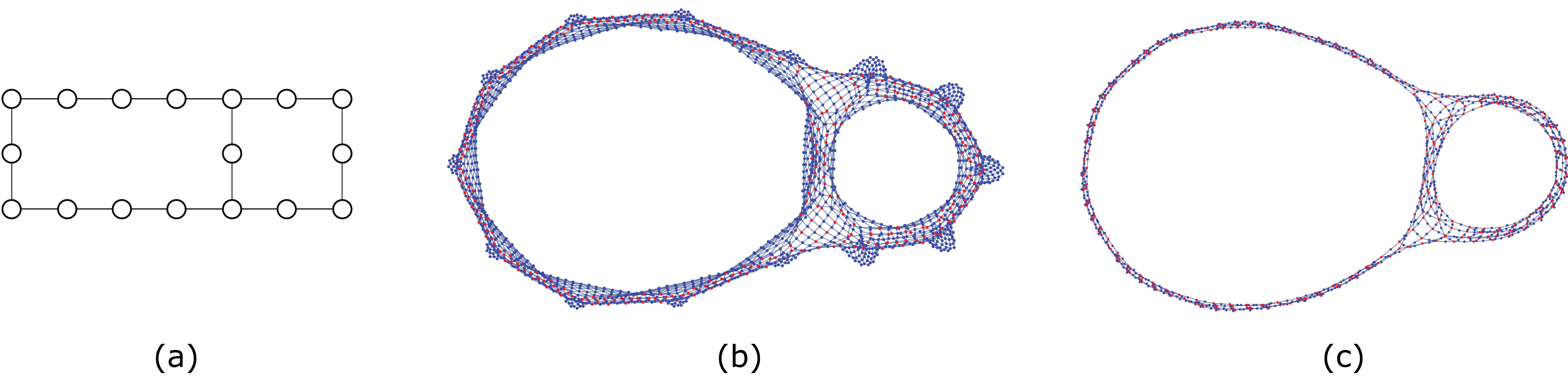}
	\protect\caption{(a) Cropped version of the Lieb lattice with three plaquettes. The site in the middle of the rung between the first two plaquettes was dropped. (b) Network representation of the single localized state of the Hamiltonian for the lattice in (a) with two holes and one flipped spin (in a ferromagnetic background). In (c) we show a cropped version of the network in (b) in which the blue nodes that are not connected to red nodes are dropped. This network is very alike to the network in (b), and this indicates that the localized state found, despite being  compact, uses a considerable part of the network. Figure generated in Mathematica 8, https://www.wolfram.com/.
	}
	\label{fig:drop}
\end{figure}

In Fig. \ref{fig:drop} (a), it is shown a cropped version of the Lieb-$1\times3$ ladder, with the site in the middle of the rung between the first two plaquettes being dropped. The network associated to the Hamiltonian for that lattice with two holes and one flipped spin (in a ferromagnetic background) is shown in Fig. \ref{fig:drop} (b), where the set of amplitudes given corresponds to the single localized state (out of eighteen of the original Hamiltonian). This state uses substantial fraction of the network. The form of this state becomes clearer when one drops the nodes of the network with zero particle wavefunction amplitude that are not connected to red nodes and finds that the cropped network obtained this way [see Fig. \ref{fig:drop} (c)] is not much different from the whole network.

\subsection{Lanczos algorithm}\label{lanczos}

Some many-body compact  localized states require a considerable fraction of the network and can not be obtained by using the methods described in the Section \ref{section6}, and full diagonalization of the corresponding Hamiltonians is too time consuming. In order to determine such states we suggest the application of a Lanczos algorithm \cite{lanczos1950iteration} after using the previous methods. This algorithm is very efficient when applied to sparse Hamiltonian matrices, and in the cases presented in this paper, they are definitively sparse since we consider systems with short-range tight-binding terms.

Next, we explain how to apply this method to determine hole localized states of the TB Hamiltonian for the Lieb lattice (or diamond chain), in the infinite coupling limit ($U/t=\infty$), that use a considerable part of the corresponding networks. In this case, the localized states are not the states with the lowest energy, but since they are zero energy state, they are the lowest energy states of the squared Hamiltonian.
Note however that in order to lift possible degeneracies associated with extended states one has first to apply local potentials (for example, to sites A in the case of the Lieb lattice and diamond chain). Thus, the energy of the localized states is no longer equal to zero, and in order to shift that energy to zero one has to add a potential (diagonal) term to the Hamiltonian.
After the previous steps, a Lanczos method should  be applied to  the Hamiltonian matrix squared in order to obtain its ground state.
Since we are interested to determine the localized states that use a substantial fraction of the network (CL states that have support in small network regions can be obtained by the methods previously described), one needs to add an energy cost to any finite projection of our state onto the already determined CL states (to increase their energy). However, if there are more than one localized state that uses a considerable fraction of the network, the Lanczos algorithm will only select one of them. In this case, one may use the Jacobi-Davidson algorithm \cite{sleijpen2000jacobi} which is a variational method good at dealing with degeneracies, and is still very efficient when applied to sparse Hamiltonian matrices.

\subsection{Symmetries and local spin rotation}

If a given Hamiltonian with many-particle localized states has symmetries, then, as usual, the full set of CL states may be obtained from a subset of CL states, applying these symmetries. The many-hole Hubbard Hamiltonian of the Lieb lattice in the infinite coupling limit ($U/t=\infty$) admits localized states such that the holes are localized in small regions of the lattice, making possible to treat the rest of the lattice as an independent system, and since this rest is half-filled, it has a local SU(2) spin symmetry.\cite{essler1991complete,ostlund1991local,carmelo2010global} On the other hand, the compact regions where holes localize have the usual global SU(2) spin symmetry of the Hubbard model. This implies that given a CL state, many others can be found applying any combination of the local spin-raising and spin-lowering operators at any site of the rest of the lattice and of spin-raising and spin-lowering operator defined within the region where the CL particle has support (a Lieb plaquette in the case of the Lieb lattice).
So, as mentioned in the Introduction, it is possible to determine part of localized states subspace relying on symmetry arguments.

Note that one can add interactions to the Hubbard Hamiltonian that change its symmetries and one may still have localized states. For instance, one may add a nearest-neighbor interacting term to the infinite coupling Hubbard model on the Lieb lattice and compact one-hole  localized states will still exist. 
This results from the fact that a hole CL state in the Lieb lattice does not involve components with different number of nearest-neighbor pairs of particles. Also one may break the SU(2) symmetry adding a Zeeman term (which only cares about the number of particles with up and down spin) and again compact one-hole localized states will still exist. 

\subsection{Equal amplitudes}\label{eqamplitudes}

In the case of small systems, one may still use the full diagonalization of the Hamiltonian in order to determine subspaces of localized states, since it may have a computing time of the order of the previous methods. However, the usual process of diagonalizing the Hamiltonian to obtain eigenstates may not give the most CL states, giving instead eigenstates that are a linear combination of the most CL states. In the Introduction, we mention that when Fourier transforming the flat band Bloch states in order to obtain maximally localized Wannier states, one may generate  the most compact form of the localized states. The relative phases of the Bloch states have to be adjusted with that goal in mind and furthermore this method does not work if the most CL states are not orthogonal. 
Here, we present an easier procedure that allows one to obtain CL states without having to first determine the maximally localized Wannier states of \textbf{real} flat-band Hamiltonians. The method works in the following way. First, one finds the eigenvalues and eigenstates as usual (direct diagonalization of the Hamiltonian), and identifies the localized states, $\psi_{loc}$, and the respective energies, $\epsilon_{loc}$. Then, one uses the $\epsilon_{loc}$ to determine again localized states, using the eigenvalue equation
\begin{equation}\label{eigenvalues}
\big[H \big]\big[ \psi_{loc}' \big]= \epsilon_{loc} \big[ \psi_{loc}' \big].
\end{equation} 
but assuming that the finite amplitudes of the localized states, $\psi_{loc}'$, are all equal in absolute value. This will avoid the linear combination of degenerate CL states that have overlap.
The result of Eq. \ref{eigenvalues} is a set of CL states linearly independent from each other (or a linear combination of CL states with a clear separation of the  finite amplitude regions). This method should be applied after using the previous methods.

\section{Conclusion} 
The physics of flat bands is currently an hot topic in condensed matter studies. The experimental realization of flat bands has been achieved in, artificial, optical \cite{bercioux2009massless,goldman2011topological,wu2007flat,wu2008p}, photonic \cite{endo2010tight,mukherjee2015observation,vicencio2015observation,xia2016demonstration,zong2016observation} and metallic \cite{kajiwara2016observation,nakata2012observation} geometrically frustrated lattices. Detecting flat bands becomes more challenging in the case of real materials \cite{chen2018prediction,jiang2019dichotomy,su2018prediction}, but recently flat bands have been observed by angle-resolved photoemission spectroscopy experiments in bilayer graphene \cite{marchenko2018extremely}, and in the case of twisted bilayer graphene, under electrostatic doping, zero-resistance states (associated to a flat band), were found. \cite{cao2018unconventional}

Flat bands are usually associated to subspaces of \textbf{one-body} localized states. In this paper we addressed \textbf{many-body} CL states.
We presented methods of construction of interacting Hamiltonians such that the subspaces of CL states of the respective non-interacting TB Hamiltonians are preserved, diminished or expanded. We showed that localized states survive the addition of interactions if those interactions are projected out of the subspace of localized states. The terms that result from this projection combine and negate the effect of the original interacting terms in the subspace of localized states.
Other modifications to the network may lead to diminished or expanded subspaces of localized states. We showed that the latter can be achieved by constructing Hamiltonians that mix states with different fillings (by adding links that connect networks associated to Hamiltonians with different particle number).
Numerical methods for the determination of many-body localized states were discussed. Since most of these states are compact in the many-body network, the full diagonalization of the Hamiltonian is not necessary and larger cluster and  larger number of particles can be addressed.

\section*{Acknowledgments}
This work is funded by the FEDER funds through the COMPETE 2020 Programme and National Funds throught FCT - Portuguese Foundation for Science and Technology under the project UID/CTM/50025/2013 and under the project PTDC/FIS-MAC/29291/2017. F. D. R. Santos acknowledges the financial support from the FCT through the grant PD/BD/135009/2017. R. G. Dias appreciates the support by the Beijing CSRC.

\bibliographystyle{model1-num-names}
\bibliography{biblio}

\section*{Supplementary Material}

In this appendix, we discuss how to define  projector operators for a generic non-orthogonal basis, following reference.\cite{soriano2014theory} Let us consider an Hilbert space of dimension $N$ and a generic normalized but non-orthogonal basis set $\{\ket{i}\}$.
The overlap beween any two states of this basis defines the overlap matrix $\braket{i|j}=S_{ij}$. If this basis is the eigenbasis of a real Hamiltonian, then it can be chosen real and that means that $\mathbf{S}$ is a real symmetric matrix. 

A generic one-body operator $\hat{\mathrm{A}}$ can be written in terms of this basis as
\begin{equation}\label{eq.9}
\hat{\mathrm{A}}=\sum_{ij} \ket{i} \tilde{A}_{ij} \bra{j},
\end{equation}
where $\tilde{A}$ is the "nuclear" matrix of the operator.\cite{soriano2014theory}
In the particular case of the identity operator  we may write
\begin{equation}\label{eq.7}
\hat{\mathrm{I}}=\sum_{ij} \ket{i} X_{ij} \bra{j},
\end{equation}
where $\mathbf{X}$ is an unknown matrix.
Since 
\begin{equation}
\mel{m}{\hat{\mathrm{I}}}{n}=\braket{m|n}=  S_{mn}
\end{equation}
and 
\begin{equation}
\mel{m}{\hat{\mathrm{I}}}{n}= \bra{m}\left(\sum_{ij} \ket{i} X_{ij} \bra{j}\right)\ket{n}=\sum_{ij} S_{mi} X_{ij} S_{jn},
\end{equation}
this  implies the matrix equation  $\mathbf{S} \cdot \mathbf{X} \cdot \mathbf{S} = \mathbf{S}$ and consequently  $\mathbf{X} = \mathbf{S}^{-1}$, that is,
\begin{equation}\label{eq.7}
\hat{\mathrm{I}}=\sum_{ij} \ket{i} S_{ij}^{-1} \bra{j}
\end{equation}
(for simplicity we use the notation $S_{ij}^{-1}$ for the elements of the inverse matrix $(\mathrm{S}^{-1})_{ij}$).

The matrix elements of a generic one-body operator $\hat{\mathrm{A}}$ are easily obtained
\begin{equation}
\bra{m} \hat{\mathrm{A}} \ket{n}=\sum_{ij} \braket{m|i} \tilde{A}_{ij} \braket{j|n}= \sum_{ij} S_{mi} \tilde{A}_{ij} S_{jn}= A_{mn},
\end{equation}
and therefore  the "nuclear" matrix of the operator is $\tilde{A}= S^{-1} A S^{-1}$.

For a general basis set with a overlap matrix $\mathbf{S}$  there exist a basis set dual to this (direct) with the property $\braket{i|j^*} = \delta_{ij}$. Only for orthogonal basis sets both direct and dual sets coincide.
The dual representation of a generic operator $\hat{\mathrm{A}}$ does not give the matrix elements, instead it gives the matrix elements of the nuclear matrix.
\begin{equation}
\mel{k^*}{\hat{\mathrm{A}}}{l^*}=\bra{k^*}\left(\sum_{ij}\ket{i}\tilde{A}_{ij}\bra{j}\right)\ket{l^*}=\sum_{ij} \delta_{ki} \tilde{A}_{ij} \delta_{jl} = \tilde{A}_{kl}.
\end{equation}
In order to  obtain the matrices elements $A_{ij}$ in the direct representation, $\hat{\mathrm{A}}$ can be written in term of the dual basis sets
\begin{equation}\label{eq.13}
\hat{\mathrm{A}}=\sum_{ij} \ket{i^*} A_{ij} \bra{j^*}.
\end{equation}
The definition for $\hat{\mathrm{A}}$ written in the direct basis in Eq. \ref{eq.9} can be recovered from Eq. \ref{eq.13}, by using the  expressions
\begin{equation}\label{eq.14}
\ket{i}=\sum_{j}S_{ij}\ket{j^*},
\end{equation}
and
\begin{equation}\label{eq.15}
\ket{i^*}=\sum_{j}S_{ij}^{-1}\ket{j}.
\end{equation}
Furthermore, since $\mel{i}{\hat{\mathrm{I}}}{j}= S_{ij}$ the identity operator can be written as 
\begin{equation}
\hat{\mathrm{I}}=\sum_{ij}\ket{i^*}S_{ij}\bra{j^*}=\sum_i \ket{i^*}\bra{i}=\sum_i \ket{i}\bra{i^*}.
\end{equation}

Let us now describe a case directly related to the topic addressed in this paper. Assume that in the non-orthogonal basis $\{\ket{i}\}$, only a few states $ \{\ket{\neg \perp,m} \}$ have non-zero overlap and we wish to define a projection operator into the subspace defined by these states. This means that  $\{\ket{i}\} = \{\ket{\perp,n} \} \cup \{\ket{\neg \perp,m} \}$
and a general state can be written as
\begin{equation}
\ket{\psi}=\sum_{i} \psi_i \ket{i} =\sum_{n} \psi_{n}^\perp \ket{\perp,n} + \sum_{m} \psi_{m}^{\neg \perp} \ket{\neg \perp,m}.
\end{equation}

We will start by defining the operator $\hat{\mathrm{P}}_i$ such that 
\begin{equation}\label{eq.17}
\hat{\mathrm{P}}_i \sum_{j} \psi_j \ket{j} = \psi_i \ket{i} 
\end{equation}
It is easy to show that 
\begin{equation}\label{eq.17}
\begin{split}
\hat{\mathrm{P}}_i &= \ket{i}\bra{i^*}\\
&=\sum_j \ket{i} S_{ij}^{-1} \bra{j}.
\end{split}
\end{equation}
Note that this operator is not a projection operator according to the usual definition (projection operators are unitary and Hermitian).
Nonetheless, the usual Hermitian projection operator $\ket{i}\bra{i}$ can be obtained by using both $\hat{\mathrm{P}}_i $ and its adjoint $\hat{\mathrm{P}}^{\dagger}_i $   in the following way
$
\hat{\mathrm{P}}_i \hat{\mathrm{P}}^{\dagger}_i  =\ket{i}\bra{i}.
$

Projecting onto the  subspace of basis $ \{\ket{\neg \perp,m} \}$  is possible through the use of the  generalized operators
\begin{equation}
\hat{\mathrm{P}}_{\neg \perp}= \sum_{m,i} \ket{\neg \perp,m} S_{mi}^{-1}\bra{i} = \sum_{m} \ket{\neg \perp, m} \bra{\neg \perp, m^*}
\end{equation}
and 
\begin{equation}
\hat{\mathrm{P}}^{\dagger}_{\neg \perp}= \sum_{m,i} \ket{i} S_{im}^{-1}\bra{\neg \perp,m} = \sum_{m} \ket{\neg \perp,m^*} \bra{\neg \perp,m},
\end{equation}
that allow a  projection of bras and kets onto the non-orthogonal subspace
\begin{equation}
\hat{\mathrm{P}}_{\neg \perp} \sum_{i} \psi_i \ket{i}= \sum_{m} \psi_{m}^{\neg \perp} \ket{\neg \perp,m},
\end{equation}
\begin{equation}
\bra{\psi} \hat{\mathrm{P}}^{\dagger}_{\neg \perp}= \sum_{m} ( \psi_{m}^{\neg \perp} )^* \bra{\neg \perp,m}.
\end{equation}
The restriction of a operator to the non-orthogonal subspace is therefore obtained using both projectors in the following way
\begin{equation}
\hat{\mathrm{A}}_{\neg \perp} \equiv \hat{\mathrm{P}}^{\dagger}_{\neg \perp} \hat{\mathrm{A}} \hat{\mathrm{P}}_{\neg \perp} .
\end{equation}
Note that 
\begin{equation}
\begin{split}
\hat{\mathrm{I}}=\sum_i \ket{i}\bra{i^*} &= \sum_{n} \ket{ \perp, n} \bra{ \perp, n^*} + \sum_{m} \ket{\neg \perp, m} \bra{\neg \perp, m^*} \\
&= \hat{\mathrm{P}}_{\perp} + \hat{\mathrm{P}}_{\neg \perp},
\end{split}
\end{equation}
therefore
\begin{equation}
\hat{\mathrm{A}}_{\perp} \equiv \hat{\mathrm{P}}^{\dagger}_{ \perp} \hat{\mathrm{A}} \hat{\mathrm{P}}_{ \perp} 
\equiv (\hat{\mathrm{I}}-\hat{\mathrm{P}}^{\dagger}_{\neg \perp}) \hat{\mathrm{A}} (\hat{\mathrm{I}}-\hat{\mathrm{P}}_{\neg \perp}).
\end{equation}

\section*{Author contributions statement}
F. D. R. Santos and R. G. Dias contributed to every stage of the work, including preparing the figures and writing and reviewing the manuscript. 

\section*{Additional information}
The authors declare no competing interests.

\end{document}